\documentclass{article}
\pdfoutput=1
\usepackage[noend]{algorithm2e}
\usepackage[hidelinks]{hyperref}
\usepackage{amsmath}
\usepackage{amsfonts}
\usepackage{amssymb}
\usepackage{amsthm}
\usepackage[margin=1cm]{caption}
\usepackage{graphicx}
\usepackage{subcaption}
\usepackage{enumerate}
\usepackage[hmargin=1.25in,vmargin=1.25in]{geometry}
\theoremstyle{definition}
\newtheorem{thm}{Theorem}
\newtheorem{lem}[thm]{Lemma}

\newtheorem{defn}[thm]{Definition}
\newtheorem{rem}[thm]{Remark}

\def\QED{\begin{footnotesize} \textbf{QED} \end{footnotesize}}

\def\wolog{\textsc{wlog}}
\def\E-R{Erd$\ddot{\text{o}}$s-R\'{e}nyi}
\def\bqed{\hfill \ensuremath{\blacksquare}}
\def\proof{\noindent \textbf{Proof.  }}
\newenvironment{myproof}{\proof \nolinebreak}{\hfill \QED \\}

\title{On the Complexity of a Matching Problem with\\ Asymmetric Weights}
\author{Lily Briggs \\ \begin{normalsize} Department of Computer Science \end{normalsize}\\ \begin{normalsize} Rensselaer Polytechnic Institute \end{normalsize} \\ \begin{normalsize} Troy, NY 12180 \end{normalsize} }
\date{}

\begin{document}
\maketitle

\abstract{

We present complexity results regarding a matching-type problem related to structural controllability of dynamical systems modelled on graphs.
Controllability of a dynamical system is the ability to choose certain inputs in order to drive the system from any given state to any desired state; a graph is said to be structurally controllable if it represents the structure of a controllable system.
We define the Orientation Control Matching problem (OCM) to be the problem of orienting an undirected graph in a manner that maximizes its structural controllability.  A generalized version, the Asymmetric Orientation Control Matching problem (AOCM), allows for asymmetric weights on the possible directions of each edge.  These problems are closely related to 2-matchings, disjoint path covers, and disjoint cycle covers.  We prove using reductions that OCM is polynomially solvable, while AOCM is much harder; we show that it is NP-complete as well as APX-hard.

}

\section{Introduction}

We introduce a new matching problem that arises out of the study of structural controllability of dynamical systems.

A dynamical system is said to be controllable if it can be driven - through choice of inputs - from any initial state to any desired state \cite{Engelberg2005Mathematical}.  A linear dynamical system can be modelled on a directed network, with edges representing influences and edge weights representing the strength of the influences.  A system is said to be structurally controllable if there exists a set of edge weights on its corresponding network that results in a controllable system \cite{Lin1974Structural}.  Intuitively, structural controllability describes a property of the structure of a system, independent of exact numerical relationships.

In the context of linear, continuous-time, finite-dimensional, time-independent dynamics, Liu et al. proved that the number of inputs necessary to make a system structurally controllable is equal to the maximum of one and the number of nodes left unmatched in a maximum control matching (to be defined) \cite{Liu2011Controllability}.  Thus, finding a maximum control matching on a network reveals how controllable the corresponding dynamical system is.

The concept of a control matching was formulated by Liu et al. in \cite{Liu2011Controllability}, though they did not use this term.  Throughout this document, unless specified otherwise (e.g. ``1-matching''), we will use ``matching'' to refer to a control matching, for brevity's sake.

\begin{defn} \label{def:match} (%Definition 8 in 
From \cite{Liu2011Controllability})
In a directed graph $ G = (V, A) $, a \textbf{control matching} is a set of edges $ M \subseteq A $ such that each node in $ V $ is the head of at most one edge in $ M $ and the tail of at most one edge in $ M $.  A node is \textbf{matched} with respect to $ M $ if it is the head of an edge in $ M $, and otherwise it is \textbf{unmatched}.  The size of a matching is $ |M| $, which is also the number of matched nodes.  A \textbf{perfect control matching} is a control matching in which every node in $ V $ is matched (then $ |M| = |V| $).
\end{defn}

\begin{rem}
Since a control matching in a directed graph corresponds to a 1-matching in its bipartite representation, it is polynomially solvable to find an optimal control matching on a directed graph \cite{Liu2011Controllability}.
\end{rem}

We say a network is easy to control if it requires few inputs to be structurally controllable.  The problem we study here is to find an orientation of an undirected graph that is easiest to control.  Equivalently, our problem is to choose an orientation of an undirected graph that allows the best control matching.  
Lv-Lin et al. introduced this problem in \cite{LvLin2012Controllability}, and presented an efficient heuristic approximation algorithm for it.  We call this problem the Orientation Control Matching problem (OCM).

We formulated the Asymmetric Orientation Control Matching problem (AOCM) to generalize OCM to allow for an asymmetric weight function.  OCM and AOCM are formally defined in Definition \ref{def:cmps}.

\begin{defn} \label{def:cmps}
Let $ G = (V, E) $ be an undirected graph with an associated weight function $ w : \{(u,v),(v,u) | (u,v) \in E\} \rightarrow \mathbb{R} $, where the weight $ w(u,v) $ corresponds to the benefit of choosing to direct the edge $ (u,v) $ from $ u $ to $ v $ in an orientation of the graph.  For each orientation of $ G $, there exists a maximum control matching, where the weight of the matching is the sum of the weights corresponding to the directions of the edges in the matching.  The task of the \textbf{Asymmetric Orientation Control Matching problem (AOCM)} is to find an orientation of $ G $ with the maximum weighted control matching.  We call the uniformly-weighted version the \textbf{Orientation Control Matching problem (OCM)}.

An instance $ (G = (V, E), w) $ of AOCM can be represented as a symmetric directed graph $ G' = (V, A)$ with the same weight function, where edges $ (u,v) $ and $ (v, u) \in A $ iff $ (u,v) \in E $.  An orientation of $ G $ is isomorphic to a subgraph of $ G' $ that includes exactly one of $ (u, v) $ and $ (v, u) $ for each adjacent $ u, v \in V $; we will call such a subgraph an \textbf{orientation-set} in what follows, for brevity. %  The value of the optimal solution is the value of the maximum control matching with no 2-cycles in $ G' $.
When speaking in terms of $ G' $, it is clear to see that an optimal solution to AOCM is an orientation-set that maximizes the largest possible control matching.
\end{defn}

This paper is organized as follows.  Section \ref{sec:ocm} briefly discusses the complexity of OCM, relating it to 2-matchings and showing that it is polynomially solvable.  In Section \ref{sec:npc} we prove AOCM is NP-complete using a reduction from a cycle cover problem, and in Section \ref{sec:apx} we prove it is APX-hard using an L-reduction from Maximum Independent Set on cubic graphs.  We conclude in Section \ref{sec:con}.

\section{Complexity of OCM} \label{sec:ocm}
When all the weights are uniform, the problem reduces to finding a maximum simple 2-matching in an undirected graph.  A simple 2-matching is a set of edges such that every node is incident on at most two edges in the matching \cite{Schrijver2004Combinatorial}; it can be interpreted as a set of node-disjoint paths and cycles that contain all the nodes in the graph.
If we have a maximum simple 2-matching, we can orient the paths and cycles it consists of to be directed paths and cycles, and then orient the rest of the edges arbitrarily.  Then we will have an oriented graph with the maximum possible maximum control matching.

The problem of finding a maximum simple 2-matching was proved to be polynomial by J. Edmonds \cite{Edmonds1965Maximum, Schrijver2004Combinatorial}.  The proof involves the description of a convex polyhedron whose extreme points correspond to matchings.

\section{NP-Completeness of AOCM} \label{sec:npc}

\begin{lem} \label{lem:np}
AOCM is in NP.
\end{lem}

\proof AOCM reduces to Weighted Independent Set.  Given a symmetric, directed, weighted graph $ G = (V, A) $ representing an instance of AOCM, we can construct an undirected graph $ H $ as follows.  For each edge in $ G $, let there be a node in $ H $ corresponding to it, with the same weight.  Then for each pair of symmetric edges in $ G $, let the corresponding nodes of $ H $ be adjacent.  In addition, let two nodes in $ H $ be adjacent if the corresponding edges in $ G $ share a tail node or a head node (that is, a node is the tail of both, or the head of both).

Now a maximum-weight independent set of nodes in $ H $ corresponds to a set of edges in $ G $ with the same weight, for which no node is the head of more than one edge, no node is the tail of more than one edge, and no two symmetric edges are in the set.  This set is a control matching on a valid orientation-set of $ G $, by definition.  Likewise, a maximum-weight control matching on an optimal orientation-set on $ G $ corresponds to an independent set on $ H $ with the same weight. \bqed

\begin{defn}
\textbf{3-cycle cover problem:} Given a graph, determine whether there exists a 3-cycle cover, which is a set of vertex-disjoint simple cycles, all with at least 3 vertices, that covers all the vertices in the graph.  In directed graphs, this is abbreviated as \textbf{3-DCC}  \cite{Blaser2001Computing}.
\end{defn}

\begin{lem} \label{lem:nph}
AOCM is NP-hard.
\end{lem}

\proof By reduction from 3-DCC, which is NP-complete \cite{Blaser2001Computing,Garey1979Computers}.

\begin{figure}
\centering
\begin{subfigure}[b]{0.25\textwidth}
\centering
\includegraphics[scale=0.7]{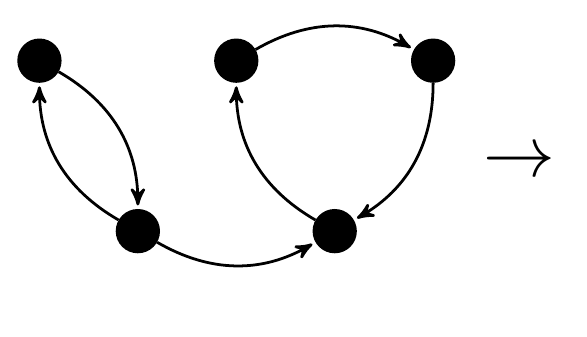}
\caption{original instance of 3-DCC \\ {}}
\end{subfigure}
\begin{subfigure}[b]{0.25\textwidth}
\centering
\includegraphics[scale=0.7]{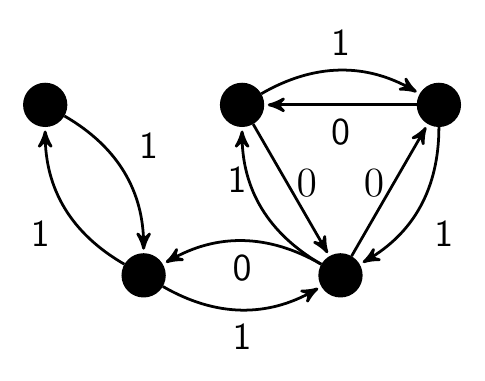}
\caption{instance of AOCM \\ {\hfill} \\ {\hfill}}
\end{subfigure}
\begin{subfigure}[b]{0.25\textwidth}
\centering
\includegraphics[scale=0.7]{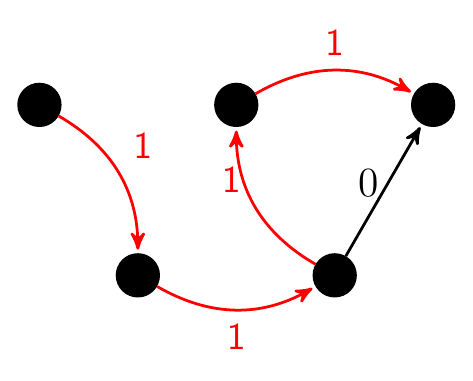}
\caption{optimal orientation of AOCM instance, with matching shown in red}
\end{subfigure}
\caption{An example of the reduction from 3-DCC to AOCM.}
\label{fig:reduc}
\end{figure}

Reduction: Let $ G = (V, A) $ be an instance of 3-DCC, and assume \wolog{} that $ G $ is simple, since self-loops and parallel edges are irrelevant to a partitioning into simple cycles larger than 3.  We can construct $ G' = (V', A') $ to be a symmetric directed graph representing an instance of AOCM; let $ V' = V $, and let $ A' $ include all the edges of $ A $.  Add edges to $ A' $ as necessary to make it symmetric.  Let $ w(u,v) = 1 $ if $ (u,v) \in A $ and 0 otherwise.  Now $ G' $ has an orientation-set that admits a control matching of weight $ |V| $ if and only if $ G $ has a 3-cycle cover.  Figure \ref{fig:reduc} illustrates the reduction.

Proof of $ \Rightarrow $:  Suppose $ G' $ has an orientation-set that admits a control matching of weight $ |V| $.  Then the matching must consist of $ |V| $ edges, all of weight 1; this means all the edges in the matching correspond to edges in the original graph $ G $.  With this many edges, the matching must make up simple cycles that cover all the vertices.  Moreover, since there are no 2-cycles in an orientation-set as defined in Definition \ref{def:cmps}, all these cycles must be of size at least 3.  Thus the corresponding edges in $ G $ make up a 3-cycle cover of $ G $.

Proof of $ \Leftarrow $: Suppose $ G $ has a directed 3-cycle cover.  Then the corresponding edges (of which there must be $ |V| $) in $ G' $ form a directed 3-cycle cover of weight $ |V| $.  If we include all of these edges in an orientation-set on $ G' $, then no matter how we choose the other edges in the orientation-set it will admit a control matching of weight $ |V| $, namely, the edges in the cycle cover. \bqed

\begin{thm}
AOCM is NP-complete. % \sp
\end{thm}

\begin{myproof}
This follows from Lemmas \ref{lem:np} and \ref{lem:nph}.
\end{myproof}

\section{APX-Hardness of AOCM} \label{sec:apx}
APX is the class of problems for which a constant-factor polynomial-time approximation algorithm exists \cite{Papadimitriou1988Optimization}.  Problems that allow PTASs, arbitrary-factor polynomial approximation algorithms, are also in APX.  Unless P=NP, though, a problem that is APX-hard does not have a PTAS; thus, Theorem \ref{thm:apx} implies that there exists a constant factor within which AOCM cannot be approximated in polynomial time (unless P=NP).

\begin{thm} \label{thm:apx}
AOCM is APX-hard. %  Moreover, it is NP-hard to approximate AOCM to within a factor of $ 140/139 - \epsilon $. 
%\sp
\end{thm}

\begin{myproof}
By L-reduction from Max-E3-Ind-Set, which is Maximum Independent Set on cubic graphs.  This has been proved to be APX-hard \cite{Alimonti1997Hardness,Berman1995Approximation}.  The method of our reduction is very similar to that in \cite{Blaser2005Approximating}.

The following two equations (from the definition of L-reduction \cite{Papadimitriou1988Optimization}) are what we will prove:

\begin{equation} \label{eq:opts}
OPT_{AOCM}(f(x)) \le \alpha \cdot OPT_{IS}(x) 
\end{equation}

\begin{equation} \label{eq:diffs}
|OPT_{IS}(x) - v(g(y))| \le \beta|OPT_{AOCM}(f(x)) - v(y)|
\end{equation}

First we will define the mapping $ f $.  Given an undirected cubic graph $ G = (V, E) $, we construct a symmetric directed weighted graph $ H = (W, A) $ which will represent an instance of AOCM.  For each edge $ (u, v) \in G $, we create two unique nodes in $ W $, and connect them with a pair of symmetric arcs; we associate one of these arcs with $ u $ and the other with $ v $.  We will use the term \textbf{edge-arcs} for arcs in $ A $ of this type.  Let $ t_{uv} $ designate the destination node of the $ v $-associate edge-arc, and $ t_{vu} $ be the destination of the $ u $-associate edge-arc.  For each node $ u \in V $, then, we will have three such pairs of nodes in $ H $.  We add two arcs, $ (t_{uv_1}, t_{v_2u}) $ and $ (t_{uv_2}, t_{v_3u}) $, to link these, where $ v_i $ ranges over the vertices adjacent to $ u $ in some arbitrary order.  These arcs are also associated with node $ u $, and will be termed \textbf{node-arcs}.  Notice that every node in $ W $ is the destination of exactly one edge-arc, and every edge-arc is associated with exactly one node in the original vertex set $V$; in addition, every node-arc associated with some $ u \in W $ is directed from the source of a $ u $-associate edge-arc to the destination of another $ u $-associate edge-arc, and no other node-arcs share the same source or destination.  One consequence of this is that no parallel edges will be created in our construction of $ H $.
The arcs described thus far are given weight 1; we then add arcs of weight 0 as necessary to make the graph symmetric.  Now $ H $ is a symmetric, directed, weighted graph, and we define $ f(G) = H$.  Due to the arbitrary element in the creation of the node-arcs, $ H $ is not necessarily unique; that is, there are multiple graphs that could be $ f(G) $, and we are defining $ f(G) $ to be one such graph, arbitrarily chosen.  The properties relevant to the proof are invariant.  %  We will also define $ H_U = (W, A') $ to be the undirected graph with associated direction-weights that $ H $ represents as an instance of AOCM, where an undirected edge $ (u, v) $ is in $ A' $ iff $ (u,v) $ (and $ (v,u) $) are in $ A $.  We do this so that we can use  
See Figures \ref{fig:chain1} and \ref{fig:k4} for examples.

\begin{figure} %[H]
	\centering
	\begin{subfigure}[b]{0.35\textwidth}
		\centering
		\includegraphics[scale = 0.8]{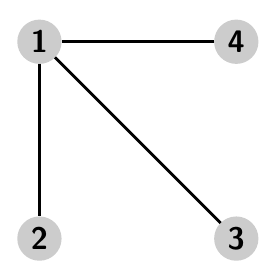}
		%\caption{}
		%\label{}
	\end{subfigure}
	\begin{subfigure}[b]{0.5\textwidth}
		\centering
		\includegraphics[scale = 0.8]{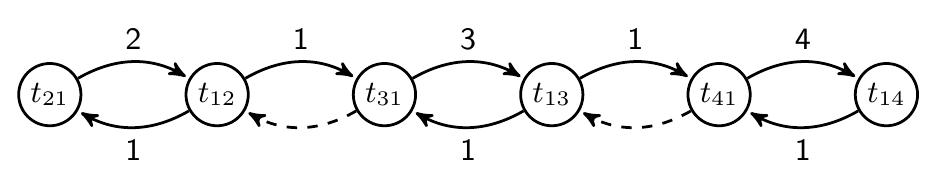}
		%\caption{}
		%\label{}
	\end{subfigure}
	\caption{This shows the construction of the portion of $ H $ (on the right) corresponding to one node and its edges in $ G $ (on the left).  The arcs are labelled with the node with which they are associated, and dashed arcs indicate arcs of weight zero.}
	\label{fig:chain1}
\end{figure}
\begin{figure}%[H]
	\centering
	\begin{subfigure}[b]{0.35\textwidth}
		\centering
		\includegraphics[scale = 0.8]{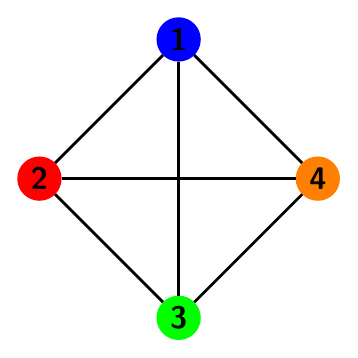}
		%\caption{}
		%\label{}
	\end{subfigure}
	\begin{subfigure}[b]{0.5\textwidth}
		\centering
		\includegraphics[scale = 0.8]{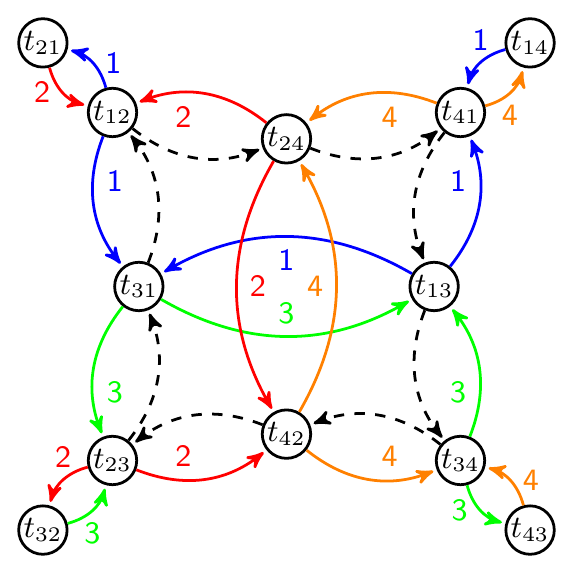}
		%\caption{}
		%\label{}
	\end{subfigure}
	\caption{On the left, $ G = K_4 $.  On the right, $ f(K_4) $ is shown with edges colored according to which node in $ G $ they are associated with.  Dashed arcs indicate arcs of weight zero.}
	\label{fig:k4}
\end{figure}

We will now describe $ g $, a function mapping feasible AOCM solutions in $ H $ to feasible Independent Set solutions in $ G $.  Given an orientation-set $ \sigma $ of $ H $, call its maximum matching $ M(\sigma) $.  Let $ S $ be the set of nodes in $ G $ whose three associated edge-arcs in $ H $ are all in $ M(\sigma) $.  We define $ g(\sigma) = S $.  To show $ S $ is an independent set, let $ u $ and $ v $ be in $ S $ and suppose they are adjacent.  Then there exists a pair of symmetric edge-arcs in $ H $, one of which is associated with $ u $ and the other with $ v $.  But if $ u $ and $ v $ are in $ S $, all of the edge-arcs associated with them must be in $ M(\sigma) $.  Since $ \sigma $ is an orientation-set and hence contains no 2-cycles, it is not possible for two symmetric arcs to be in $ \sigma $, let alone in a matching on it, so $ u $ and $ v $ must not be adjacent.

Now we will develop some properties of feasible and optimal matchings that will be useful in proving (\ref{eq:opts}) and (\ref{eq:diffs}).

\newpage
Let  $ \sigma $ be an orientation-set of $H $.  The value of $ \sigma $, $ v(\sigma) $, is the value of the maximum matching in it, $ v(M(\sigma)) $.  Since each non-zero arc in $ A $ is associated with exactly one vertex in $ V $, we can say
 \[ v(\sigma) = v(M(\sigma)) = \sum_{u \in V} (\text{\# $ u $-associate arcs in $ M $}).  \]

We can find an upper bound on the number of arcs in $ M(\sigma) $ by bounding the number of $ u $-associate arcs in $ M(\sigma) $ for each $ u \in V $.  Given a $ u \in V $, there are five arcs associated with it in $ A $: three edge-arcs, and two node-arcs.  There are four possible numbers of edge-arcs that could be in $ M $: 0, 1, 2, or 3.  We will consider each of these four cases separately; the ideas are summarized in Figure \ref{fig:cases}.  In each case, we will show that the number of $ u $-associate arcs in $ M(\sigma) $ is bounded by a certain number, based on the number of $ u $-associate \textit{edge} arcs in $ M(\sigma) $.

Case $ a $:  There are 0 edge-arcs associated with $ u $ in $ M(\sigma) $.  Consider the node-arcs associated with $ u $; the only edges they share source nodes with are edge-arcs also associated with $ u $, and same with the edges they share destination nodes with.  So, when none of the $ u $-associate edge-arcs are in $ M $, $ M(\sigma) $ can contain up to 2 (both) of the $ u $-associate node-arcs.  Note that in a sub-optimal orientation-set, one or both of the node-arcs might not be in the orientation-set; in an optimal orientation-set however, both will be, as they add to the maximum possible matching.

Case $ b $:  There is 1 edge-arc associated with $ u $ in $ M(\sigma) $.  Case $ b(i) $: If this arc is the central arc in the chain of $ u $-associate arcs (see Figure \ref{fig:cases}), it conflicts with both $ u $-associate node-arcs and thus neither of them can be in the matching.  Case $ b(ii) $: If it is one of the outer arcs in the chain, then the node-arc not sharing any endpoints with it could be in the matching.  In an optimal orientation-set, Case $ b(i) $ will never arise, as the one edge-arc could be exchanged for both node-arcs with a gain in value.  As in Case $ a $, an optimal orientation-set exhibiting Case $ b(ii) $ will always include the other node-arc, allowing it to be in the matching, while a sub-optimal orientation-set may not.

Case $ c $:  There are 2 edge-arcs associated with $ u $ in $ M(\sigma) $.  No matter where in the chain the edge-arcs are, it is impossible for either node-arc to be in $ M $.

Case $ d $:  There are 3 edge-arcs associated with $ u $ in $ M(\sigma) $.  It is again impossible for either node-arc to be in $ M $.

\begin{figure}
\centering
\begin{subfigure}[b]{\textwidth}
\centering
\includegraphics[scale=1]{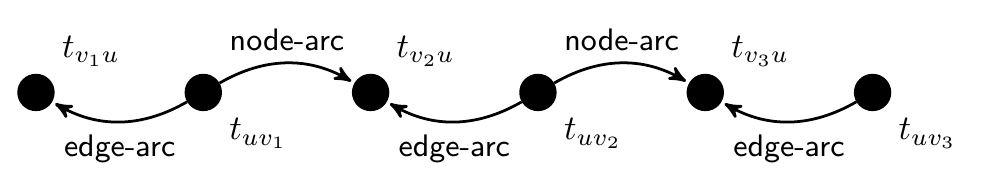}
\caption{}
\label{fig:base}
\end{subfigure} \\
\begin{subfigure}[b]{\textwidth}
\centering
\[\begin{array}{|c|c|c|c|c|}
\hline
A	&	B	&	C	&	D	&	E \\
\hline
a	&	0	&	\includegraphics[scale=1]{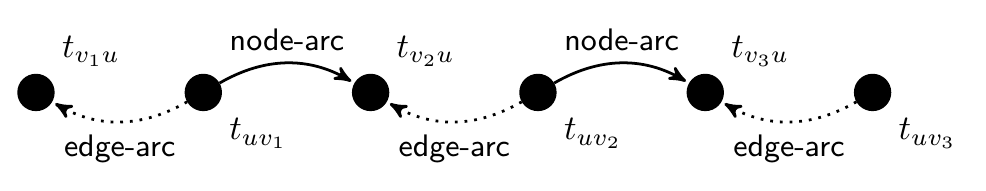}	&	0,1,2	&	2 \\
\hline
b	&	1	&	\includegraphics[scale=1]{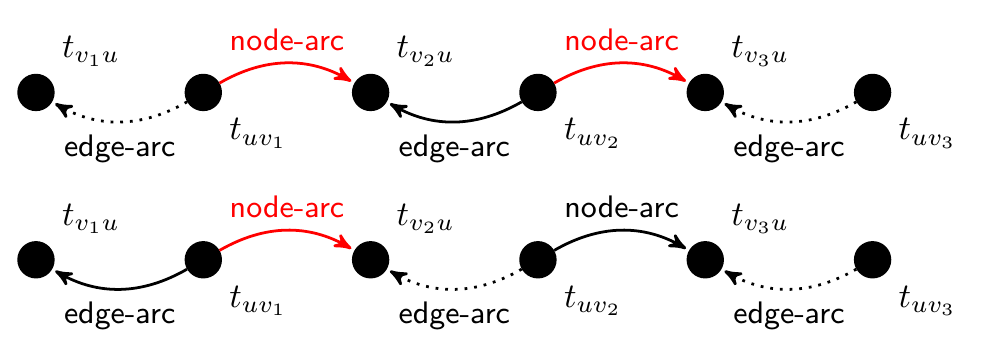}	&	0,1	&	2 \\
\hline
c	&	2	&	\includegraphics[scale=1]{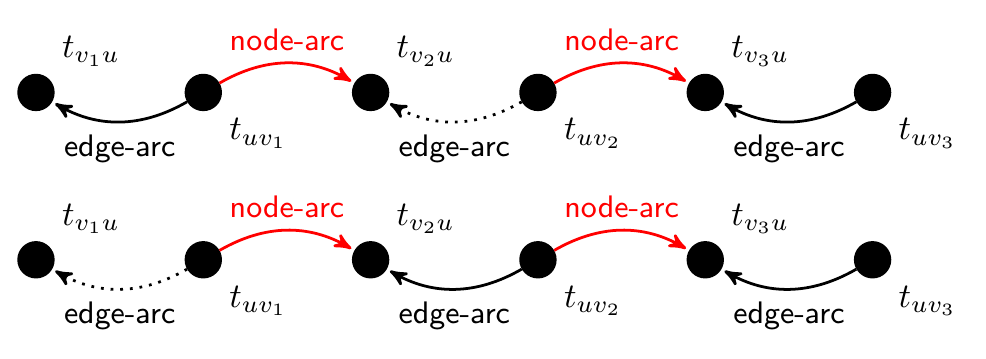}	&	0	&	2 \\
\hline
d	&	3	&	\includegraphics[scale=1]{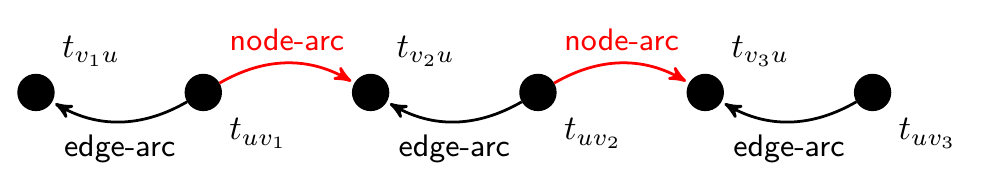}	&	0	&	3 \\
\hline
\end{array}\]
\caption{}
\label{fig:table}
\end{subfigure}

\caption{Figure \ref{fig:base} depicts the edge- and node-arcs associated with a given vertex $ u $.  These are the only arcs relevant to an analysis of the four cases.  Fig. \ref{fig:table} illustrates the different cases. Key: $ A $: Case letter. $ B $: \# edge-arcs associated with $ u $ in $ M $. $ C $: Possible configurations.  $ D $: \# node-arcs associated with $ u $ possibly in $ M $.  $ E $: Optimal total \# of arcs associated with $ u $ in $ M $.  The diagrams represent the following.  The only arcs drawn are those associated with $ u $.  Solid edge-arcs are those we assume to be in $ M $, while dotted edge-arcs we assume to not be in $ M $.  Red node-arcs cannot be in $ M $ given what we have assumed about the edge-arcs, and black node-arcs could feasibly be in $ M $.}

\label{fig:cases}
\end{figure}

Given any matching, we can partition the nodes of $ V $ into 4 sets: $ V_0 $, $ V_1 $, $ V_2 $, and $ V_3 $, where $ V_i $ is the set of nodes with $ i $ associated edge-arcs in the matching.  We will use superscripts to designate the orientation-set whose maximum matching induces a particular partition; e.g., $ V_i^y $ is the set of nodes with $ i $ associated edge-arcs in the maximum matching on an orientation-set $ y $.  So \[ v(y) = \sum_{i=0}^3\sum_{u \in V_i} (\text{\# u-associate arcs in M(y)}).  \]

\begin{lem} \label{lem:values}
Let $ n $ be the number of nodes in a given cubic graph $ G = (V,E) $.  Let $ H = f(G) $, and let $ y $ be an orientation-set of $ H $.  Then $ v(y) \le 2n + |V_3^y| $.  Let $ \sigma^* $ be an optimal orientation-set of $ H $; then $ v(\sigma^*) = 2n + |V_3^{\sigma^*}| $.
\end{lem}

\proof  From looking at the four cases, we know that 
\[ v(y) = \sum_{u \in V_0^y}(\text{0,1 or 2}) + \sum_{u \in V_1^y}(\text{1 or 2}) + \sum_{u \in V_2^y} 2 + \sum_{u \in V_3^y} 3. \]
Thus 
\[ v(y) \le \sum_{u \in \bigcup_{i=0}^2 V_i^y}2 + \sum_{u \in V_3^y} 3 = 2|\bigcup_{i=0}^3 V_i^y| + |V_3^y|.\]
Since $ \bigcup_{i=0}^3 V_i^y = V$, and $|V| = n $, $ v(y) \le 2n + 2|V_3^y| $.

In an optimal orientation-set $ \sigma^* $, we know from looking at the four cases that 
\[ v(\sigma^*) = \sum_{u \in \bigcup_{i=0}^2 V_i^{\sigma^*}}2 +  \sum_{u \in V_3^{\sigma^*}}3. \]
It follows, similarly to above, that $ v(\sigma^*) = 2n + 2|V_3^{\sigma^*}| $. \bqed \\

%\[ v(M(\sigma^*)) = \sum_{u \in V_0^{\sigma*}}2 +  \sum_{u \in V_1^{\sigma*}}2 + \sum_{u \in V_2^{\sigma*}}2 + \sum_{u \in V_3^{\sigma*}}3 \]
An upper bound on $ v(\sigma^*) $ can be derived by noting that $ 2n + 2|V_3^{\sigma^*}| $ is maximized when $ V_3^{\sigma^*} = V $, with a value of $ 3n $.  This is clearly never attainable, but allows us to prove condition \ref{eq:opts}.  Since any cubic graph has an independent set of size at least $ \frac{n}{4} $ \cite{Greenlaw1995Cubic}, $ OPT_{AOCM} (f(G)) \le 3n \le 12\cdot OPT_{IS}(G) $ .

Now for condition \ref{eq:diffs}.  We will first prove another lemma.

\begin{lem} \label{lem:opts}
Let $ G = (V,E) $ be an instance of Independent Set, let $ H = f(G) $, and let $ \sigma^* $ be an optimal orientation-set on $ H $.  Then $ v(g(\sigma^*)) = OPT_{IS}(G) $.
\end{lem} 

\proof Let $ G $, $ H $, and $ \sigma^* $ be as defined in Lemma \ref{lem:opts}.  Let $ y^* $ be a solution to 
\[ max\{v(y): \text{y is an orientation-set of $ H $, }v(g(y)) = OPT_{IS}(G)\}. \]

From Lemma \ref{lem:values}, we know that $ v(y^*) \le 2n + |V_3^{y^*}| $.  According to our analysis of cases $ a $ and $ b $, for any $ u \in V $ we can always orient the node-arcs associated with $ u $ in a way that will allow 2 $ u $-associate arcs to be in the maximum matching.  So since $ v(y^*) \ge v(y) $ for all $ y $ s.t. $ |V_3^y| = |V_3^{y^*}| $,  \begin{equation} \label{eq:ystar}
v(y^*) = 2n + |V_3^{y^*}|.
\end{equation}

In addition, since $ |V_3^{y^*}| = |g(y^*)| = OPT_{IS}(G) $, for any orientation-set $ \bar{y} $ of $ H $ we have $ |V_3^{y^*}| \ge |V_3^{\bar{y}}| $.  Specifically, \begin{equation} \label{eq:yvsigma}
 |V_3^{y^*}| \ge |V_3^{\sigma^*}|.
\end{equation}

Putting (\ref{eq:ystar}) and (\ref{eq:yvsigma}) together along with Lemma \ref{lem:values}, we have $ v(y^*) \ge v(\sigma^*) $.  Thus $ v(y^*) = v(\sigma^*) $, which implies that $ |V_3^{\sigma^*}| = |V_3^{y^*}| $, so then $ v(g(\sigma^*)) = v(g(y^*)) = OPT_{IS}(G) $. \bqed \\

From here, it is not difficult to show condition \ref{eq:diffs}:

\[ \begin{array}{rcll} \vspace{8pt}
OPT_{AOCM}(f(G)) - v(y) & \ge & 2n + |V_3^{\sigma^*}| - (2n + |V_3^y|) & \text{(Lemma \ref{lem:values})} \\
						&  =  & |V_3^{\sigma^*}| - |V_3^y| & \\
						&  =  & v(g(\sigma^*)) - v(g(y)) & \\
						&  =  & OPT_{IS}(G) - v(g(y)) & \text{(Lemma \ref{lem:opts})}
\end{array}
\]

%It remains to prove the bound of $ 140/139 - \epsilon $.  It has been shown that Maximum Independent Set on cubic graphs cannot be approximated to within a factor of $ 140/139 - \epsilon $ for all $ \epsilon > 0 $ \cite{Berman1999Some}.  Since we have shown the existence of functions $ f $ and $ g $ such that $ OPT_{IS}(G) - v(g(y)) \le OPT_{AOCM}(f(G)) - v(y) $, it must be that AOCM has at least as tight a bound on its approximability as $ 140/139 - \epsilon $.

\end{myproof}

\section{Conclusion} \label{sec:con}
We have discussed a matching problem, OCM, and introduced a variation on it, AOCM.
We have derived several complexity results for these problems, proving that OCM is polynomially solvable while AOCM is NP-complete and APX-hard.

The identification of OCM as being simply a different formulation of maximum 2-matching suggests numerous approaches for solving or approximating it efficiently, since 2-matchings have been studied extensively \cite{Araoz2007Simple,Edmonds1970Matching,Letchford2008Odd, Padberg1982Odd}.

The difficulty classification of AOCM informs potential work on approximation algorithms.  It may be fruitful to adapt algorithms for related problems, such as 2-matchings, weighted path or cycle covers, or ATSP \cite{Blaser2005Approximating,Blaser2001Computing, Vishwanathan1992Approximation}.  In addition, since AOCM reduces to Maximum Weighted Independent Set, algorithms for that problem could also be used to approximate AOCM.

\section*{Acknowledgements}
The author thanks Dr. Krishnamoorthy %, Dr. Anshelevich, and Dr. Mitchell for helpful advice.  The author also thanks 
for helpful advice, and both Dr. Krishnamoorthy and James Hampton for critiquing the manuscript.  The author was supported by a Teaching Assistantship at Rensselaer Polytechnic Institute.

\bibliographystyle{abbrv}
\bibliography{../orient}

\begin{thebibliography}{10}

\bibitem{Alimonti1997Hardness}
P.~Alimonti and V.~Kann.
\newblock {Hardness of approximating problems on cubic graphs}.
\newblock In G.~Bongiovanni, D.~P. Bovet, and G.~Battista, editors, {\em
  Algorithms and Complexity}, volume 1203 of {\em Lecture Notes in Computer
  Science}, chapter~26, pages 288--298. Springer, Berlin, 1997.

\bibitem{Araoz2007Simple}
J.~Ar\'{a}oz, E.~Fern\'{a}ndez, and O.~Meza.
\newblock {A simple exact separation algorithm for 2-matching inequalities}.
\newblock Technical report, Technical University of Catalonia, Barcelona,
  Spain, Nov. 2007.

\bibitem{Berman1995Approximation}
P.~Berman and T.~Fujito.
\newblock {On approximation properties of the Independent set problem for
  degree 3 graphs}.
\newblock In S.~Akl, F.~Dehne, J.-R. Sack, and N.~Santoro, editors, {\em
  Algorithms and Data Structures}, volume 955 of {\em Lecture Notes in Computer
  Science}, pages 449--460. Springer, Berlin, 1995.

\bibitem{Blaser2005Approximating}
M.~Bl\"{a}ser and B.~Manthey.
\newblock {Approximating Maximum Weight Cycle Covers in Directed Graphs with
  Weights Zero and One}.
\newblock {\em Algorithmica}, 42(2):121--139, Apr. 2005.

\bibitem{Blaser2001Computing}
M.~Bl\"{a}ser and B.~Siebert.
\newblock {Computing Cycle Covers without Short Cycles}.
\newblock In F.~Heide, editor, {\em Algorithms — ESA 2001}, volume 2161 of
  {\em Lecture Notes in Computer Science}, pages 368--379. Springer, 2001.

\bibitem{Edmonds1965Maximum}
J.~Edmonds.
\newblock {Maximum matching and a polyhedron with 0,1-vertices}.
\newblock {\em Journal of Research of the National Bureau of Standards: Section
  B Mathematics and Mathematical Physics}, 69B(1-2):125--130, 1965.

\bibitem{Edmonds1970Matching}
J.~Edmonds and E.~L. Johnson.
\newblock {Matching: a well-solved class of integer linear programs}.
\newblock In {\em in: Combinatorial structures and their applications (Gordon
  and Breach}, pages 89--92, 1970.

\bibitem{Engelberg2005Mathematical}
S.~Engelberg.
\newblock {\em {A Mathematical Introduction to Control Theory}}.
\newblock Series in Electrical and Computer Engineering. Imperial College
  Press, London, 2005.

\bibitem{Garey1979Computers}
M.~R. Garey and D.~S. Johnson.
\newblock {\em {Computers and Intractability: A Guide to the Theory of
  NP-Completeness}}.
\newblock W. H. Freeman \& Co., New York, NY, USA, 1979.

\bibitem{Greenlaw1995Cubic}
R.~Greenlaw and R.~Petreschi.
\newblock {Cubic graphs}.
\newblock {\em ACM Comput. Surv.}, 27(4):471--495, Dec. 1995.

\bibitem{Letchford2008Odd}
A.~N. Letchford, G.~Reinelt, and D.~O. Theis.
\newblock {Odd Minimum Cut Sets and $b$-Matchings Revisited}.
\newblock {\em SIAM J. Discret. Math.}, 22(4):1480--1487, Oct. 2008.

\bibitem{Lin1974Structural}
C.-T. Lin.
\newblock {Structural controllability}.
\newblock {\em Automatic Control, IEEE Transactions on}, 19(3):201--208, June
  1974.

\bibitem{Liu2011Controllability}
Y.-Y. Liu, J.-J. Slotine, and A.-L. Barab\'{a}si.
\newblock {Controllability of complex networks.}
\newblock {\em Nature}, 473(7346):167--173, May 2011.

\bibitem{LvLin2012Controllability}
H.~Lv-Lin, L.~Song-Yang, L.~Gang, and B.~Liang.
\newblock {Controllability and Directionality in Complex Networks}.
\newblock {\em Chinese Physics Letters}, 29(10):108901+, Oct. 2012.

\bibitem{Padberg1982Odd}
M.~W. Padberg and M.~R. Rao.
\newblock {Odd Minimum Cut-Sets and b-Matchings}.
\newblock {\em Mathematics of Operations Research}, 7(1):67--80, Feb. 1982.

\bibitem{Papadimitriou1988Optimization}
C.~Papadimitriou and M.~Yannakakis.
\newblock {Optimization, approximation, and complexity classes}.
\newblock In {\em Proceedings of the twentieth annual ACM symposium on Theory
  of computing}, STOC '88, pages 229--234, New York, NY, USA, 1988. ACM.

\bibitem{Schrijver2004Combinatorial}
A.~Schrijver.
\newblock {\em {Combinatorial Optimization : Polyhedra and Efficiency
  (Algorithms and Combinatorics)}}.
\newblock Springer, July 2004.

\bibitem{Vishwanathan1992Approximation}
S.~Vishwanathan.
\newblock {An approximation algorithm for the asymmetric travelling salesman
  problem with distances one and two}.
\newblock {\em Information Processing Letters}, 44(6):297--302, Dec. 1992.

\end{thebibliography}

\end{document}